\documentclass[runningheads]{llncs}

\usepackage[T1]{fontenc}
\usepackage[utf8]{inputenc}
\usepackage{lmodern}
\usepackage{amsmath,amssymb,amsfonts,mathtools}
\usepackage{booktabs,tabularx,array}
\usepackage{enumitem}
\usepackage{float}
\usepackage{placeins}
\floatstyle{ruled}
\floatname{algorithm}{Algorithm}
\usepackage[ruled,vlined,linesnumbered]{algorithm2e}
\usepackage{microtype}
\usepackage{xurl}
\usepackage{xcolor}
\usepackage{comment}
\usepackage{hyperref}
\usepackage{cleveref}

\hypersetup{hidelinks}

\AddToHook{env/proof/end}{\qed}

\crefname{theorem}{Theorem}{Theorems}
\crefname{lemma}{Lemma}{Lemmas}
\crefname{proposition}{Proposition}{Propositions}
\crefname{corollary}{Corollary}{Corollaries}
\crefname{definition}{Definition}{Definitions}
\crefname{conjecture}{Conjecture}{Conjectures}
\crefname{remark}{Remark}{Remarks}
\crefname{problem}{Open Problem}{Open Problem}

\newcommand{\FindExchange}{\textnormal{\textsc{FindExchange}}}
\newcommand{\Decompose}{\textnormal{\textsc{ComponentDecomposition}}}

\definecolor{rr_color}{named}{black}

\title{Matroid Connectedness and Component Decomposition: Breaking the Quadratic Barrier}
\titlerunning{Matroid Connectedness and Component Decomposition}
\author{Manuel Aprile, Roberto Ronco}
\authorrunning{M. Aprile\inst{1}\orcidID{0000-0002-6805-6903}, 
R. Ronco\inst{2}\orcidID{0000-0001-6846-6374}}
\institute{
Department of Mathematics, University of Padua, Via Trieste 63,
I-35121, Padua, Italy \\
\email{manuel.aprile@unipd.it}
\and
Institute of Marine Engineering, Italian National Research Council, Via De Marini 6, 
I-16149, Genoa, Italy \\
\email{roberto.ronco@cnr.it}
}

\begin{document}
\maketitle

\begin{abstract}
    Deciding whether a matroid is connected and computing its connected components are two fundamental algorithmic problems in matroid theory. Since the 1970s, the best known algorithms for both problems have used $O(n^2)$ independence queries on an $n$-element ground set. Our main contribution is to break this quadratic barrier: we give a deterministic algorithm that computes all connected components of a matroid on $n$ elements with $O(n \log{n})$ independence queries. Our algorithm for component decomposition combines fundamental basis exchanges with a disjoint-set union data structure, avoiding the explicit construction of the fundamental graph. Moreover, we show that our algorithm is asymptotically optimal by giving a lower bound based on a simple information-theoretic argument: any (deterministic or bounded-error randomized) algorithm to compute connected components needs $\Omega(n \log{n})$ independence queries. Then, we turn to the query complexity of deciding whether a matroid is connected. For this problem, we present an algorithm that uses $O(nh)$ independence queries, where $h \ge 1$ bounds the size of any chain of nonempty proper cyclic flats. The algorithm achieves asymptotically optimal linear query complexity on elementary split matroids, that include all paving matroids, and on arbitrary direct sums of uniform matroids.
\end{abstract}

%\medskip

%\noindent\textbf{Mathematics Subject Classification.}
%05B35, % Combinatorial aspects of matroids and geometric lattices
%68Q25, % Analysis of algorithms and problem complexity
%68W40. % Analysis of algorithms

\section{Introduction}\label{sec:intro}

Connectedness is one of the classical structural notions in matroid theory~\cite{Tutte1966,Cunningham1981,CunninghamEdmonds1980}. Two basic related problems are deciding whether a matroid is connected and computing its connected components. The algorithmic approaches to these problems depend on the \emph{oracle model} used to access the given matroid.
A well-known example is the \emph{rank oracle}, which, given a matroid $M = (E, \mathcal{I})$ on $n = |E|$ elements, returns the rank $r_M(S)$ of a subset $S \subseteq E$. Relatedly, \emph{CUT oracles} have been extensively studied for graphs and hypergraphs~\cite{RubinsteinSchrammWeinberg2018,ApersEtAl2022,ChakrabartyLiao2026}. 
For a hypergraph $\mathcal{H} = (V, \mathcal{F})$, a CUT query on $S \subseteq V$ returns the number of hyperedges having nonempty intersection with both $S$ and $V \setminus S$. Chakrabarty and Liao~\cite{ChakrabartyLiao2026} recently showed that the connected components of an arbitrary hypergraph on $n$ vertices can be recovered using $O(n)$ expected CUT queries. 
This, in particular, applies to computing matroid components, through the concept of circuit hypergraph and the corresponding circuit-CUT queries (which we discuss further at the end of Section~\ref{sec:decomp}).

The focus of this article is the \emph{independence oracle}. In this model, the matroid %collection of independent sets 
is accessible only through queries that decide whether a subset $S \subseteq E$ belongs to $\mathcal{I}$. 
Rank queries are at least as informative as independence queries, since $S$ is independent if and only if $r_M(S) = |S|$; conversely, $r_M(S)$ can be recovered using at most $|S|$ independence queries by greedily constructing a maximal independent subset of $S$. Independence, rank, and circuit-CUT queries therefore expose different amounts of information, and these differences are central to the oracle complexity of matroid connectedness.

The seminal work of Cunningham~\cite{Cunningham1973} gives a deterministic $O(n^2)$-query algorithm for deciding whether a matroid is connected in the independence-oracle model.
Later, Krogdahl~\cite{Krogdahl1977} studied the dependence graph associated with a basis and showed that its connected components coincide with the matroid components, thereby also yielding an $O(n^2)$-query algorithm for computing them. Given a basis $B$, this \emph{fundamental graph} is the bipartite graph with parts $B$ and $E\setminus B$, where $u\in B$ and $v\in E\setminus B$ are adjacent if and only if $B-u+v$ is a basis. 
Explicitly testing all $r(n-r)$ possible fundamental exchanges of a rank-$r$ matroid again gives a quadratic algorithm, an approach also used in recent work~\cite{CohenAddadEtAl2026}.
Bixby and Cunningham~\cite{BixbyCunningham1979} gave a polynomial independence-oracle algorithm for testing 3-connectivity.
%based on a recursive characterization through separating cocircuits.
More recently, Huang, Feng, and Li~\cite{HuangFengLi2026} studied the query complexity of deciding ordinary matroid connectedness in both the classical and quantum settings, in the independence oracle model. 
In particular, Huang, Feng, and Li claim a $\Omega(n^2)$ lower bound for the problem of deciding whether a matroid is connected. However, the argument has a fatal flaw, as we argue in Section~\ref{sec:decomp}, where we give a counterexample to the basis-removal property used in their proof.
In the following, we refer to:
\begin{itemize}[itemsep=4pt, label=$\circ$]
    \item \emph{Problem (A)}, as the problem of deciding whether a matroid is connected, and
    \item \emph{Problem (B)}, as the problem of computing all the connected components of a matroid,
\end{itemize} 
both in the independence oracle model.

\smallskip

Our first contribution settles the worst-case complexity of Problem (B): 
%we give an $O(n \log{n})$ algorithm in the independence oracle model for computing all the connected components of a matroid 
we give a deterministic algorithm that computes the connected components of a matroid using $O(n\log n)$ independence queries
and prove an $\Omega(n \log{n})$ lower bound for both deterministic and bounded-error randomized algorithms. The upper bound combines a \emph{Disjoint Set Union} (DSU) data structure~\cite{CormenEtAl2022} with the $\FindExchange$ algorithm presented in~\cite{ChakrabartyEtAl2019}, allowing fundamental exchanges to be located without explicitly constructing the entire fundamental graph.

Our second contribution concerns Problem (A). 
We give an $O(nh)$-query algorithm for a matroid $M$ whose chains of nonempty proper \emph{cyclic flats}, i.e., the flats of $M$ that are union of circuits, are bounded by $h \ge 1$.
We show that this algorithm uses $O(n)$ independence queries on \emph{elementary split matroids}~\cite{BercziEtAl2023} and on arbitrary direct sums of uniform matroids.
%If $h = 0$, the algorithm uses $O(n)$ independence queries.

Elementary split matroids contain all paving matroids~\cite{BercziEtAl2023} and thus, under the conjecture that asymptotically almost all matroids are paving~\cite{MayhewEtAl2011}, this algorithm makes a linear number of independence queries on asymptotically almost all matroids.
This result may help isolate the structure of the instances on which a possible superlinear lower bound for Problem (A) must rely. 
More generally, strong oracle lower bounds are known for other problems related to matroid connectedness. Jensen and Korte~\cite{JensenKorte1982} showed that computing the exact connectivity parameter, namely the minimum value of $k$ for which the matroid admits a $k$-separation, requires a superpolynomial number of independence-oracle queries.
For matroid intersection, Harvey~\cite{harvey2008matroid}
proved that deterministic algorithms need $(\log_2 3)n-o(n)$ independence
queries, and Blikstad, Mukhopadhyay, Nanongkai and Tu~\cite{BlikstadEtAl2023}
improved this to $\Omega(n\log n)$; the argument from~\cite{BlikstadEtAl2023} reduces from the two-party
communication complexity of graph connectivity.
%the randomized case remains open.
Whether bounded-error randomized algorithms for matroid intersection
require a superlinear number of independence queries remains open.

The remainder of the article is organized as follows. 
In Section~\ref{sec:prelim}, we introduce the notation, definitions, and known results used throughout the article. 
In Section~\ref{sec:decomp}, we present the $O(n\log{n})$-query algorithm and matching lower bound for computing the connected components of a matroid. 
In Section~\ref{sec:conn}, we give an algorithm for deciding matroid connectedness, and show that it is linear in some special cases.

\section{Preliminaries}\label{sec:prelim}

In this section, we recall notation and results that will prove useful in the remainder of the paper.

% \cite{Oxley2011}
Let $M = (E, \mathcal{I})$ denote a \emph{matroid} having rank function $r_M$. 
If $B$ is a basis of $M$ and $e \in E \setminus B$, then $B \cup \{e\}$ contains a unique circuit $C(e, B)$, which is called the \emph{fundamental circuit} of $e$ with respect to $B$. If $e$ is a loop, we set $C(e, B) = \{e\}$. We denote by $U_{k,n}$ the \emph{uniform matroid} on $n$ elements where every set of at most $k$ elements is independent.
Hereafter, whenever convenient for notation purposes, we shall denote the set difference and union operators by $-$ and $+$ instead of $\setminus$ and $\cup$, respectively.

Given matroids $M$, $M_1$, and $M_2$, we say that $M$ is the \emph{direct sum} (or 1-sum) of $M_1$ and $M_2$, and refer to it as $M = M_1 \oplus M_2$, 
if the ground set of $M$ is the disjoint union of the ground sets of $M_1$ and $M_2$, and a set $I$ is independent in $M$ if and only if 
$I = I_1 \cup I_2$ where $I_i$ is independent in $M_i$, for $i \in \{1, 2\}$.
A matroid is \emph{connected} if it cannot be written as the direct sum of two matroids, each with non-empty ground set, and disconnected otherwise. The direct sum is commutative and associative, and every non-empty matroid $M$ can be written as a direct sum of (one, two or possibly more) connected matroids in a unique way, up to the order of the summands; the ground sets of these matroids are the \emph{connected components} of $M$. Equivalently, two elements $x$ and $y$ lie in the same connected component of $M$ if $x = y$ or there exists a circuit of $M$ containing both.

An \emph{independence query} on $M$ takes any subset $A \subseteq E$ and determines whether $A$ is independent in $M$. A deterministic algorithm using independence queries on $M$ is associated with a rooted binary decision tree induced by all the queries and the related sequential branches; we refer to the maximum height of such tree as the \emph{query complexity} of the associated algorithm.

\smallskip

Let $B$ denote a basis of $M$. We refer to the \emph{fundamental graph} of $M$ with respect to $B$ as the bipartite graph $G_B(M) = (B ~ \dot\cup ~ (E \setminus B), F_B)$ such that edge $\{b, e\} \in F_B$ if and only if $B - b + e \in \mathcal{B}(M)$, 
which in turn is equivalent to $b \in C(e, B)$. 
% \lambda_M(S) = r_M(S) + r_M(E \setminus S) - r_M(E).
Krogdahl~\cite{Krogdahl1977} proved that the connected components of $G_B(M)$ coincide with those of $M$: 
\begin{theorem}\label{thm:krogdahl}
  For every basis $B$, the graph-connected components of $G_B(M)$ are exactly the connected components of $M$.
\end{theorem}
%two elements of $E$ belong to the same matroid component if and only if they are connected by a path in $G_B$.
In particular, $M$ is connected if and only if $G_B(M)$ is connected. Thus, the component partition induced by $G_B$ is independent of the chosen basis $B$.
Given a basis $B$, a straightforward algorithm constructs $G_B(M)$ by testing whether $B - b + e$ is independent for each pair $(b, e) \in B \times (E \setminus B)$. 
This requires exactly $r(n - r)$ independence-oracle queries. The connected components of $M$ can then be determined by a graph search in $G_B(M)$, without further oracle queries, yielding an $O(n^2)$ query upper bound.

\smallskip

For $e \in E \setminus B$, let
\begin{equation}
  N_B(e) = C(e, B) \setminus \{e\}
\end{equation}
denote the \emph{fundamental neighborhood} of $e$ with respect to $B$. Observe that $N_B(e)$ corresponds to the neighbourhood of $e$ in the graph $G_B(M)$.
We recall the following straightforward consequence of the fundamental-circuit property~\cite[cfr. Corollary~1.2.6]{Oxley2011}, which will prove useful in the remainder of the article.
\begin{lemma}\label{lem:subset-test}
  Let $M = (E, \mathcal{I})$ denote a matroid on ground set $E$, and let $B$ denote one of its bases.
  For every $A\subseteq B$ and $e \in E \setminus B$,
  \begin{equation*}
    B - A + e \in \mathcal{I} 
    \quad \Leftrightarrow\quad
    A \cap N_B(e) \neq \varnothing.
  \end{equation*}
\end{lemma}
\begin{proof}
  The set $B + e$ contains a unique circuit, namely $C(e, B)$. Hence, $B - A + e$ is independent if and only if $A$ intersects $C(e, B)$, or equivalently, $N_B(e)$.
\end{proof}

In view of Lemma~\ref{lem:subset-test}, it takes a single independence query to check, for any $A\subseteq B$ and $e\in E\setminus B$, whether $A$ contains a neighbor of $e$ (without necessarily identifying one).
By repeatedly halving the candidate set, such queries can be used to find a neighbor through the binary-search procedure $\FindExchange$, described in Algorithm~\ref{alg:findexchange}.

\begin{algorithm}[t]
    \caption{$\FindExchange(M,B,e,A)$}\label{alg:findexchange}
    \noindent\textbf{Input:} a matroid $M = (E, \mathcal{I})$ with independence oracle access, a basis $B$ of $M$, $e \in E \setminus B$, nonempty $A\subseteq B$.
    \par
    \noindent\textbf{Output:} $b\in A$ with $B - b + e \in \mathcal{I}$, or \textsc{Null}.
        \begin{enumerate}[label=\arabic*:,leftmargin=1.7em,itemsep=0pt,parsep=0pt,topsep=3pt]
        \item \textbf{if} $B - A + e$ is dependent \textbf{then}
        \item \hspace*{1em}  \textbf{return} \textsc{Null}
        \item \textbf{while} $|A| > 1$ \textbf{do}
        \item \hspace*{1em}  split $A = A_1 \,\dot\cup\, A_2$ with $|A_i|\le \lceil |A|/2 \rceil$
        \item \hspace*{1em}  \textbf{if} $B - A_1 + e$ is independent \textbf{then}
        \item \hspace*{2em}  $A \gets A_1$
        \item \hspace*{1em}  \textbf{else}
        \item \hspace*{2em}  $A \gets A_2$
        \item \textbf{return} the unique $b\in A$
    \end{enumerate}
\end{algorithm}

\section{An optimal algorithm for component decomposition}\label{sec:decomp}

In this section, we present a deterministic $O(n\log{n})$ algorithm to compute all the matroid components of any matroid $M = (E, \mathcal{I})$ on ground set $E$ with $|E| = n$. Huang, Feng, and Li~\cite{HuangFengLi2026} claim an $\Omega(n^2)$ lower bound for the weaker problem of deciding whether $M$ is connected, even for bounded-error randomized algorithms. However, their argument is flawed, as it relies on a property (cfr. Lemma~7, \cite{HuangFengLi2026}) that does not hold in general.

More precisely, let $E = E_0 ~\dot\cup~ \overline{E}_0$, where $|E_0| = q$, and consider the connected rank-$q$ matroid $M'$ with basis set
$\mathcal{B} = \{E_0\} \cup \{E_0 - x + y : x\in E_0, y \in \overline{E}_0\}$.
The lower bound argument relies on the claim that, for every $B^* \in \mathcal{B}$, the family $\mathcal{B} \setminus \{B^*\}$ is the basis family of a disconnected matroid. While this holds for $B^* = E_0$, it fails for any $B^* = E_0 - x + y$ when both $q$ and $n - q$ are at least two, as the resulting family does not satisfy the basis-exchange axiom.
For example, consider $E_0 = \{1, 2\}$ and $\overline{E}_0 = \{3, 4\}$. Removing the basis $B^* = \{1, 3\}$ leaves the family $\mathcal{B} \setminus \{B^*\} = \{\{1, 2\}, \{1, 4\}, \{2, 3\}, \{2, 4\}\}$, which violates the basis-exchange axiom for the bases $\{2, 3\}$ and $\{1, 4\}$.
In general, the following holds:
\begin{proposition}\label{prop:counter}
    Let $E=E_0\mathbin{\dot\cup}\overline E_0$ with
    $|E_0|$ and $|\overline E_0|\ge 2$. In addition, let
    \[
    \mathcal B=\{E_0\}\cup
    \{E_0-x+y:x\in E_0,\ y\in\overline E_0\}.
    \]
    For each $x\in E_0$ and $y\in\overline E_0$,
    $\mathcal B\setminus\{E_0-x+y\}$ is not the family of bases
    of a matroid.
\end{proposition}

We report the proof of Proposition~\ref{prop:counter} in Appendix~\ref{app:prop-counterex}.

\medskip

\begin{algorithm}[!t]
    \caption{$\Decompose(M)$}\label{alg:decompose}
    \noindent\textbf{Input:} a matroid $M = (E, \mathcal{I})$ with independence oracle access.
    \par
    \noindent\textbf{Output:} the connected components of $M$.
    \begin{enumerate}[label=\arabic*:,leftmargin=1.7em,itemsep=0pt,parsep=0pt,topsep=3pt]
        \item construct an arbitrary basis $B$ of $M$ greedily
        \item initialize a DSU with one class for each element of $E$
        \item \textbf{for all} $e \in E \setminus B$ \textbf{do}
        \item \hspace*{1em} \textbf{while} true \textbf{do}
        \item \hspace*{2em} $A \gets \{b \in B : \textsc{Find}(b) \ne \textsc{Find}(e)\}$
        \item \hspace*{2em} \textbf{if} $A = \varnothing$ \textbf{then}
        \item \hspace*{3em} \textbf{break}
        \item \hspace*{2em} $b \gets \FindExchange(M, B, e, A)$
        \item \hspace*{2em} \textbf{if} $b = \textsc{Null}$ \textbf{then}
        \item \hspace*{3em} \textbf{break}
        \item \hspace*{2em} $\textsc{Union}(e, b)$
        \item \textbf{return} the DSU classes
    \end{enumerate}
\end{algorithm}

The algorithm for component decomposition, reported in Algorithm~\ref{alg:decompose}, exploits the \emph{disjoint-set union} (DSU) data structure~\cite{CormenEtAl2022}, which maintains a partition of a ground set ad supports two operations: \textsc{Find}$(x)$, which identifies the representative of the element of the partition containing $x$, and \textsc{Union}$(x, y)$, which merges the two elements of the partition containing $x$ and $y$, respectively. With \emph{path compression} and \emph{union by rank}, two well-known implementation techniques for DSU, a sequence of $m$ operations on $n$ elements requires $O((m + n)\alpha(n))$ operations, where $\alpha(n)$ is the inverse Ackermann function~\cite{CormenEtAl2022} that is upper bounded by $4$ for every $n \le 2^{2^{16}}$.

For a given input matroid $M = (E, \mathcal{I})$, Algorithm~\ref{alg:decompose} constructs a basis $B$ of $M$ greedily and initializes a DSU structure on the ground set $E$. By construction, each element in the initial partition of the structure is a singleton. 
%For each $e \in E \setminus B$, the algorithm considers the basis elements that are not in the same component of $e$, i.e., the set $A_e = \{b \in B : \textsc{Find}(b) \ne \textsc{Find}(e)\}$.
At each iteration of the inner loop, for each $e \in E \setminus B$, the algorithm forms the set 
$A_e = \{b \in B : \textsc{Find}(b) \ne \textsc{Find}(e)\}$ of basis elements outside the current DSU class of $e$.
If $\textsc{FindExchange}$ returns $b \in A_e$, then $\{e, b\}$ is an edge of the fundamental graph, and thus the algorithm performs $\textsc{Union}(e, b)$.
We stress that Algorithm~\ref{alg:decompose} does not reconstruct the whole fundamental graph, but finds a spanning forest, as we argue below: this allows us to obtain its connected components, which is all the information needed to solve the component decomposition problem.

\begin{theorem}\label{thm:comp-decomp-alg}
    Let $M = (E, \mathcal{I})$ denote a matroid with $|E| = n$, for an integer $n \ge 2$. Assume that $M$ has rank $r$ and $c$ connected components.
    Then, Algorithm~\ref{alg:decompose} returns exactly the $c$ connected components of $M$, and, for $1 \le r \le n - 1$, uses at most
    \begin{equation}\label{eq:comp-query-bound}
        n + (n - r) + (n - c) \bigl( 1 + \lceil{\log_2{r}\rceil} \bigr) 
    \end{equation}
    independence queries.
\end{theorem}
\begin{proof}
    First, observe that, for any $A\subseteq B$ and $e \in E \setminus B$ as in the algorithm, $\textsc{FindExchange}$ returns \textsc{Null} if and only if $A\cap N_B(e) = \varnothing$. Otherwise, it returns an element of $A\cap N_B(e)$ using at most $1 + \lceil \log_2{|A|} \rceil$ oracle queries.
    Every successful call to $\textsc{FindExchange}$ identifies an edge of the fundamental graph of $M$ that joins two distinct DSU classes. Conversely, when the inner loop terminates for an element $e\in E \setminus B$, all its neighbors in the fundamental graph belong to its DSU class. Hence, upon termination, the DSU classes coincide with the connected components of the fundamental graph of $M$, and therefore with the connected components of $M$.
    
    Now, constructing $B$ greedily requires at most $n$ oracle queries. For each $e\in E\setminus B$, at most one call to $\textsc{FindExchange}$ returns \textsc{Null}, requiring exactly one query. Thus, there are at most $n - r$ such calls. Moreover, each successful call results in a union of two distinct DSU classes. Since the
    number of classes decreases from $n$ to $c$, there are exactly $n-c$ successful calls, each requiring at most $1 + \lceil \log_2{r} \rceil$ oracle queries.
    Summing these contributions yields the upper bound~\eqref{eq:comp-query-bound}.
\end{proof}
Theorem~\ref{thm:comp-decomp-alg} immediately implies the following result.
\begin{corollary}
    The connectedness of a matroid on $n \ge 2$ elements can be
    decided deterministically using $O(n\log n)$ independence queries.
\end{corollary}

Finally, we show that the complexity of Algorithm~\ref{alg:decompose} is tight.

We use a simple information-theoretic argument inspired by~\cite{ChakrabartyLiao2024}, which gives a similar lower bound for the problem of learning partitions with independence queries. %We extend a proof to cover randomized algorithms and expected query complexity.

%\bigskip

%\bigskip
\begin{theorem}\label{thm:decompose-lower-bound}
    Computing the connected components of a matroid on
    $n$ elements requires $\Omega(n\log{n})$ independence oracle
    queries in the worst case. This holds for deterministic algorithms
    and randomized algorithms succeeding with probability at least
    $2/3$ on every input, including when complexity is measured by the
    worst-case expected number of queries. The lower bound holds
    already for graphic matroids. %, hence also for regular and binary matroids, even when a basis is supplied.
\end{theorem}
\begin{proof}
    Let $n=2m$ and consider the two following disjoint sets:
    \[
        B=\{b_1,\ldots,b_m\}
        \quad \text{and} \quad 
        X=\{x_1,\ldots,x_m\}.
    \]
    For each permutation $\pi\in S_m$, define
    \[
    M_\pi=\bigoplus_{i=1}^{m}
    U_{1,2}\bigl(\{b_i,x_{\pi(i)}\}\bigr),
    \]
    where $\oplus$ denotes the operation of direct sum and $U_{1,2}(\{a, b\})$ denotes the uniform matroid $U_{1,2}$ on ground set $\{a,b\}$.
    A set is independent in $M_\pi$ precisely when it contains at most one
    element from each pair $\{b_i,x_{\pi(i)}\}$. Notice that these pairs are exactly the  connected components of $M_\pi$. %Thus the $m!$ inputs have distinct component    partitions: a correct output uniquely determines $\pi$.
    We also remark that $M_\pi$ is graphic, as it can be represented by disjoint pairs of parallel
    edges. % and has $B$ as a basis. 
    Now, consider an algorithm for computing the connected components of a matroid: since the $m!$ matroids $M_\pi$ have distinct component partitions, the output of the algorithm must uniquely determine $\pi$.
    
      \medskip
    %\noindent\emph{Deterministic algorithms.}
    Suppose that the algorithm is deterministic and makes at most $q$ independence queries, hence it can be represented as a binary decision tree of depth at most $q$, with at most
    $2^q$ leaves. Each leaf specifies one output partition and can
    therefore be correct for at most one permutation.
    Correctness on every input implies
    \begin{equation*}
        2^q\ge m!
        \quad\Leftrightarrow\quad
        q\ge \left\lceil\log_2(m!)\right\rceil
        =\Omega(n\log{n})
    \end{equation*}

    \medskip
    %\noindent\emph{Randomized algorithms with a fixed query bound.}
    Suppose now that the algorithm is randomized, and correct with probability at least $\frac23$ on every input. We can see the algorithm as drawing a random tape $\rho$, independently of the input,
    and then running deterministically; let $T_\rho$ be the resulting decision tree.
    If the algorithm makes at most $q$ queries on every execution, then
    every $T_\rho$ has depth at most $q$. Let
    \[
    G_\rho=\{\pi\in S_m : T_\rho \text{ outputs the component partition of } M_\pi\}.
    \]
    By the deterministic argument applied to $T_\rho$, we have $|G_\rho|\le 2^q$ for
    every $\rho$. %: the tree has at most $2^q$ leaves, and each leaf is correct for at most one $\pi$, since the partitions of the matroids $M_\pi$ are pairwise distinct.
    The success probability on $M_\pi$ is $\Pr_\rho[\pi\in G_\rho]$. Summing over
    all permutations and using linearity of expectation,
    \[
    \sum_{\pi\in S_m}\Pr\nolimits_\rho[\pi\in G_\rho]
    =\mathbb{E}_\rho 
    \Bigl[ \sum\nolimits_{\pi\in S_m}\mathbf 1\{\pi\in G_\rho\} \Bigr]
    =\mathbb{E}_\rho\bigl[|G_\rho|\bigr]
    \le 2^q .
    \]
    On the other hand, each of the $m!$ terms on the left is at least $2/3$ by
    assumption. Hence,
    \begin{equation*}
        2^q\ge \tfrac{2}{3}\,m!
        \quad\Leftrightarrow\quad
        q\ge \log_2(m!)-\log_2 \tfrac{3}{2}=\Omega(n\log n).
    \end{equation*}

    \medskip
    %\noindent\emph{Expected query complexity.}
    Finally, consider the same randomized algorithm as above, but suppose that its number $T=T(M,\rho)$  of independence queries is a random variable depending on the input $M$ and on random tape $\rho$. Set
    \[
    C=\max_{\pi\in S_m}\mathbb{E}_\rho[T(M_\pi,\rho)].
    \]
    If $C=\infty$, the claim is immediate.
    Otherwise, truncate the algorithm after
    $L=\lceil 6C\rceil+1$ queries, declaring failure if it has not
    terminated. By Markov's inequality, on each input the probability
    of truncation is at most $C/L\le 1/6$.
    The truncated algorithm therefore succeeds with probability at
    least $2/3-1/6=1/2$ on every input.
    The preceding counting argument yields $2^L\ge m! / 2$.
    Since $L\le 6C+2$, it follows that
    \[
    C\ge \tfrac{1}{6}(\log_2(m!)-3) %\frac{\log_2(m!)-3}{6}
    =\Omega(n\log{n}).
    \]
 %   \qed
    % Here we used $\log_2(m!)=m\log_2 m-O(m)$.
    % For odd $n$, take $m=\lfloor n/2\rfloor$ and add one fixed
    % coloop to every input, including it in the supplied basis.
    % The same bounds apply.

    This concludes the proof for even $n$. For odd $n$, consider the instances obtained by adjoining a fixed coloop $z$ to each $M_\pi$ instance on $n-1$ elements. An independence query on a set $A$ is simulated by querying $A\setminus\{z\}$ in the original instance. The component $\{z\}$ is known, so the same lower bound follows. The extended instances remain graphic by representing $z$ as a bridge.
\end{proof}

We conclude the section with a consequence of Theorem \ref{thm:decompose-lower-bound} for the hypergraph CUT queries mentioned in the introduction. For a matroid $M$, let
$\mathcal{H}(M)$ be its circuit hypergraph, with vertex set $E$ and the circuits of $M$
as hyperedges. We now show that answering CUT queries on $\mathcal{H}(M)$ by using an independence oracle of $M$ requires at least a logarithmic number of calls. Note that this can be immediately seen for a single CUT query, since the answer of a CUT query can take $\Omega(n)$ distinct values, while an independence query gives a binary answer. 
However, Theorem~\ref{thm:decompose-lower-bound} shows that this cost cannot be
amortized away when making multiple queries on the same circuit hypergraph: suppose that a deterministic procedure, which may store all previous answers,
answers each CUT query on $\mathcal{H}(M_\pi)$ using at most $t$ independence queries (and makes no other independence query).
Recall that the aforementioned algorithm of~\cite{ChakrabartyLiao2026} computes the connected components of a matroid using $O(n)$ CUT queries in expectation. Combining the two algorithms on $\mathcal{H}(M_\pi)$ will recover the components of
$M_\pi$ using $O(n t)$ independence queries in expectation. Hence Theorem~\ref{thm:decompose-lower-bound} implies $t=\Omega(\log n)$.
The same conclusion holds for randomized procedures with expected cost $t$ per query. We give more details in Appendix~\ref{app:corr-cut-simulation-proof}.

%%%%%%%%%%%%%%%%%%%%%%%%%%%%%%%%%%%%%%%%%%%%%%%%%%%%%%%%%%%%
%%%%%%%%%%%%%%%%%%%%%%%%%%%%%%%%%%%%%%%%%%%%%%%%%%%%%%%%%%%%
%%%%%%%%%%%%%%%%%%%%%%%%%%%%%%%%%%%%%%%%%%%%%%%%%%%%%%%%%%%%

\section{Linear query bounds for matroid connectedness}\label{sec:conn}

In this section, we present an algorithm (Algorithm~\ref{alg:conn}) that decides matroid connectedness in the independence oracle model.
Let $M = (E, \mathcal{I})$ denote a matroid on a ground set $E$ with $n \ge 2$ elements, and let $B$ denote a basis of $M$.
Throughout the section, for a given $X \subseteq E$, we refer to $r_M(X)$ as the \emph{rank} of $X$ in $M$ and write $r = r_M(E)$ for the rank of $M$. We also refer to the closure of $X$ in $M$ as $\text{cl}_M(X)$.
In addition, a set $F \subseteq E$ is a \emph{flat} if $\text{cl}_M(F) = F$. A flat is \emph{cyclic} if it is a (not necessarily disjoint) union of circuits of $M$.
For integers $a$ and $b$, we denote the set of the integers $c$ satisfying $a \le c \le b$ by $[a, b]$.
Hereafter, we identify the elements of $M$ with the vertices of its fundamental graph $G_B$.
In addition, for brevity, we refer to a chain of nonempty proper cyclic flats simply as a chain of cyclic flats, unless specified otherwise.

%\begingroup
%\SetAlFnt{\footnotesize}
\begin{algorithm}[!t]
    \caption{\textsc{MatroidConnectedness($M$)}}
    \label{alg:conn}
    \KwIn{A matroid $M=(E,\mathcal{I})$ with $|E| \ge 2$, where $\mathcal{I}$ is given through an independence oracle.}
    \KwOut{\textsc{connected} if $M$ is connected, and \textsc{disconnected} otherwise.}
    
    $B \gets \varnothing$\; \label{alg:conn:first-B}
    \ForEach{$e \in E$}{
        \If{$B + e \in \mathcal{I}$}{
            $B \gets B + e$\; \label{alg:conn:last-B}
        }
    }
    \If{$|B| \in \{0,|E|\}$}{\label{alg:conn:if-all}
        \Return{\textsc{disconnected}}\; \label{alg:conn:ret-disc-1}
    }
    \ForEach{$e \in E - B$}{ \label{alg:conn:loop-chk-begin}
        \If{$\{e\} \not\in \mathcal{I}$}{
            \Return{\textsc{disconnected}}\;\label{alg:conn:ret-disc-2}
        }
    } \label{alg:conn:loop-chk-end}
    Choose $e_0 \in E - B$\; \label{alg:conn:choose-e0}
    %$N_B(e_0) \gets \{b \in B : (B \setminus \{b\}) \cup \{e_0\} \in \mathcal{I}\}$\;
    $S \leftarrow N_B(e_0)$\; \label{alg:conn:init-S}
    \While{$S \ne B$}{ \label{alg:conn:while-begin}
        $f \gets \textsc{Null}$\;
        \ForEach{$e \in E - B$}{ \label{alg:conn:innermostfor-begin}
            \If(\tcp*[f]{$N_B(e) \cap S \ne \varnothing$}){$B - S + e \in \mathcal{I}$}
            { \label{alg:conn:c1}
                \If(\tcp*[f]{$N_B(e) - S \ne \varnothing$}){$S + e \in \mathcal{I}$}
                { \label{alg:conn:c2}
                    $f \gets e$\;
                    \textbf{break}\;
                }
            }
        } \label{alg:conn:innermostfor-end}
        \If{$f = \textsc{Null}$}{
            \Return{\textsc{Disconnected}}\; \label{alg:conn:ret-disc-3}
        }
        %$N_B(f) \leftarrow \{ b \in B : (B \setminus \{b\}) \cup \{f\} \in \mathcal{I} \}$\; \label{alg:conn:compute-N}
        $S \leftarrow S + N_B(f)$\; \label{alg:conn:upd-S}
    } \label{alg:conn:while-end}
    \Return{\textsc{Connected}}\; \label{alg:conn:ret-conn}
\end{algorithm}
%\endgroup

Algorithm~\ref{alg:conn} starts by constructing a basis $B$ of $M$ and evaluating some sufficient conditions for disconnection, i.e., whether the elements in $E$ are all loops or all coloops, or if at least one of the elements in $E$ is a loop.
Afterwards, it explores the connected component of some chosen element $e_0 \in E - B$ in the fundamental graph $G_B$ of $M$, by maintaining a set $S \subseteq B$ of the vertices in $B$ reached so far. The algorithm initializes $S$ to the fundamental neighborhood $N_B(e_0)$, and then searches for an element $e \in E - B$ whose fundamental neighborhood has nonempty intersection with both $S$ and $B - S$.
For such an $e$, Algorithm~\ref{alg:conn} updates $S$ as $S + N_B(e)$ and then continues the search.
However, if no iteration yields such an $e$, then $M$ is disconnected because $S \ne B$, and Algorithm~\ref{alg:conn} outputs accordingly; otherwise, if $S = B$, then $M$ is connected.

The bound on the number of independence queries used by Algorithm~\ref{alg:conn} depends on the number of updates of $S$, which, as we now show, is bounded by the size of the largest chain of cyclic flats in $M$.

\begin{theorem}\label{thm:conn-alg-correct}
    Let $M = (E, \mathcal{I})$ denote a matroid on $|E| = n \ge 2$ elements with rank $r$.
    Algorithm~\ref{alg:conn} correctly decides whether $M$ is connected. 
    If all chains of nonempty proper cyclic flats in $M$ contains at most $h$ flats, for some $h \ge 1$, then Algorithm~\ref{alg:conn} uses $O(nh)$ independence queries.
\end{theorem}
\begin{proof}
    First, we prove the correctness of Algorithm~\ref{alg:conn}.
    Let $B$ denote the basis constructed by Algorithm~\ref{alg:conn} in lines \ref{alg:conn:first-B}--\ref{alg:conn:last-B}.
    If $|B| = 0$, all elements of $E$ are loops, and if $|B| = n$, all elements of $E$ are coloops. Since $n \ge 2$, either case implies that $M$ is disconnected. In addition, if $e$ is a loop, then $M$ is disconnected, and all loops of $M$ belong to $E - B$, and thus lines \ref{alg:conn:if-all}--\ref{alg:conn:loop-chk-end} are correct.
    
    Next, Algorithm~\ref{alg:conn} chooses an arbitrary nonbasis element $e_0 \in E - B$ and initializes $S$ to $N_B(e_0)$. This set is nonempty because $M$ is loopless, and each of its vertices is adjacent to $e_0$ in the fundamental graph.
    %Throughout the algorithm, $S$ contains the elements of $B$ that have already been identified as reachable from $e_0$ in the fundamental graph.
    While $S \ne B$, Algorithm~\ref{alg:conn} considers each $e \in E - B$ until it finds an element $f$ satisfying the conditions at both lines~\ref{alg:conn:c1} and \ref{alg:conn:c2}. 
    By Lemma~\ref{lem:subset-test}, for each candidate $e$, such two conditions correspond to 
    \begin{equation*}
        N_B(e) \cap S \ne \varnothing, \qquad N_B(e) \cap (B - S) = N_B(e) - S \ne \varnothing,
    \end{equation*}
    respectively.
    If such an $f$ is found, let $s \in N_B(f) \cap S$; for each $b \in N_B(f) - S$, the edges $\{s, f\}$ and $\{f, b\}$ form a path from $s$ to $b$. 
    Hence, the update at line~\ref{alg:conn:upd-S} preserves the property that each element of $S$ is reachable from $e_0$.
    As $N_B(f) - S$ is nonempty, the set $S + N_B(f)$ at line~\ref{alg:conn:upd-S} is strictly larger than $S$, and thus Algorithm~\ref{alg:conn} terminates.
    
    If $f$ is \textsc{Null} after lines \ref{alg:conn:innermostfor-begin}--\ref{alg:conn:innermostfor-end}, there are no other basis elements in the component of $e_0$ to be discovered. 
    Indeed, by contradiction, suppose that there is a path from $e_0$ to a vertex in $B - S$ after lines \ref{alg:conn:innermostfor-begin}--\ref{alg:conn:innermostfor-end}. Its first vertex in $B$ also belongs to $S$ because $N_B(e_0) \subseteq S$. As the path ends in $B - S$,
    it contains two vertices $s \in S$ and $b \in B - S$, where $s$ is the last vertex in $S$ in the path, and $b$ is the first vertex in $B$ outside $S$.
    Since $G_B$ is bipartite, there is a vertex $e \in E - B$ both adjacent to $s$ and $b$ in the path.
    However, such an $e$ would satisfy the conditions at lines~\ref{alg:conn:c1} and \ref{alg:conn:c2}, contradicting $f = \textsc{Null}$.
    If $S \ne B$, the algorithm correctly returns \textsc{Disconnected} at line~\ref{alg:conn:ret-disc-3}.

    On the other hand, if $S = B$, then each element in $B$ is reachable from $e_0$, and each element of $E - B$ has a neighbor in $S=B$ because Algorithm~\ref{alg:conn} verified that $M$ is loopless at lines \ref{alg:conn:loop-chk-begin}--\ref{alg:conn:loop-chk-end}.
    Hence, $G_B$ is connected, and Algorithm~\ref{alg:conn} correctly returns \textsc{Connected} at line~\ref{alg:conn:ret-conn}.

    \smallskip

    Next, we prove that Algorithm~\ref{alg:conn} uses $O(nh)$ independence queries. First, observe that lines~\ref{alg:conn:first-B}--\ref{alg:conn:init-S} use at most $2n$ queries.
    %Let now $p$ denote the number of iterations of lines~\ref{alg:conn:while-begin}--\ref{alg:conn:while-end} and $k$ denote the number of updates performed by line~\ref{alg:conn:upd-S}.
    Let $k$ denote the number of updates performed by line~\ref{alg:conn:upd-S}.
    If $k=0$, then Algorithm~\ref{alg:conn} executes the iterations at lines \ref{alg:conn:innermostfor-begin}--\ref{alg:conn:innermostfor-end} exactly once, and thus the bound follows since $h\geq 1$. Hence, assume that $k\geq 1$.
    %\CorrM{Observe that $k\geq 1$ because no $e \in E \setminus B$ is a loop by lines \ref{alg:conn:loop-chk-begin}--\ref{alg:conn:loop-chk-end} and $h \ge 1$ by assumption.}
    Each iteration of lines~\ref{alg:conn:while-begin}--\ref{alg:conn:while-end} either updates $S$ or returns \textsc{Disconnected}, and thus the number of iterations of lines~\ref{alg:conn:while-begin}--\ref{alg:conn:while-end} is either $k$ or $k + 1$.
    Each execution of the loop at lines~\ref{alg:conn:innermostfor-begin}--\ref{alg:conn:innermostfor-end} and the update at line~\ref{alg:conn:upd-S} use at most $2(n - r)$ and $r$ queries, respectively.
    Hence, Algorithm~\ref{alg:conn} requires at most $O(nk)$ independence queries.

    Now, we show that $k$ is bounded by $h$. Let $e_j$ denote the element selected at the $j$-the update of $S$ and let $S^{(j)}$ denote the value of $S$ after the $j$-th update, where we assume $S^{(0)} = N_B(e_0)$.
    For each $j$ satisfying $0 \le j \le k$, we have $S^{(j)} = \cup_{i = 0}^j N_B(e_i)$ by construction. 
    Since $e_i$ belongs to the closure of $N_B(e_i)$ for each $i \in [0, j]$, we also have:
    \begin{equation*}
        F_j := \text{cl}_M(S^{(j)}) 
            = \text{cl}_M \left( \bigcup_{i = 0}^j C(e_i, B) \right).
    \end{equation*}
    We claim that $F_j$ is a cyclic flat for each $j \in [0, k]$. Indeed, let $X = \bigcup_{i = 0}^j C(e_i, B)$. Then, each $e \in \text{cl}_M(X) - X$ lies in a circuit contained in $X + e$. Together with the circuits whose union is $X$, these circuits cover $\text{cl}_M(X)$, so this flat is cyclic.

    The flats $F_j$, for $j \in [0, k]$, form an inclusionwise monotonic chain. 
    %\CorrR{As $S^{(j)}$ is independent and, by definition, $r_M(\text{cl}(S^{(j)})) = r_M(S^{(j)})$, then $r_M(F_j) = |S^{(j)}|$ for each $j \in \{0, 1, \ldots, k + 1\}$.  (this sentence is not needed, right?)}
    Moreover, since the sets $S_j$ are independent and their size increases strictly, the flats $F_j$ have strictly increasing ranks. Since $F_0$ contains the nonempty set $N_B(e_0)$, we have:
    \begin{equation*}
        \varnothing \subset F_0 \subset \cdots \subset F_k \subseteq E.
    \end{equation*}
    Hence, the first $k$ flats form a chain of cyclic flats, and thus $k \le h$.
    %Finally, by assumption $p \le k+1$ holds, and thus~\eqref{eq:conn-query-count} is bounded by $O(nh)$, completing the proof. 
\end{proof}
\noindent
Observe that, if $M$ has no nonempty proper cyclic flats, i.e., $h=0$, then Algorithm~\ref{alg:conn} also decides connectedness correctly using
$O(n)$ independence queries, as it either returns \textsc{Disconnected} during the preliminary checks at lines \ref{alg:conn:if-all}--\ref{alg:conn:ret-disc-2} or satisfies $S=B$ immediately after the initialization of $S$ at line~\ref{alg:conn:init-S}.

The size of chains of cyclic flats can grow linearly with $n$ even when Algorithm~\ref{alg:conn} iterates over each nonbasis element exactly once.
For example, consider the matroid $M$ equal to the direct sum of $t$ uniform matroids $U_{1, 2}$.
For $i$ satisfying $1 \le i < t$, the unions of the ground sets of the first $i$ summands form a chain of $t - 1$ cyclic flats. 
The structure of the fundamental neighborhoods of direct sums of uniform matroids yields an upper bound on the number of iterations that is independent of the sizes of cyclic flats.
\begin{corollary}\label{cor:uni-sum-lin-cmp}
    Let $M$ denote a direct sum of $t \ge 1$ uniform matroids on $n \ge 2$ elements.
    Then, Algorithm~\ref{alg:conn} decides whether $M$ is connected or disconnected by using $O(n)$ independence queries.
\end{corollary}
We report the proof of Corollary~\ref{cor:uni-sum-lin-cmp} in Appendix~\ref{app:cor-uni-sum-proof}.

\medskip

Interestingly, we can show that Algorithm~\ref{alg:conn} uses $O(n)$ independence queries also for \emph{elementary split matroids}, introduced by Bérczi et al.~\cite{BercziEtAl2023} as a subclass of split matroids~\cite{JoswigSchroeter2017} that contains all connected split matroids and, interestingly, all paving matroids.
Elementary split matroids are exactly those matroids whose nonempty proper cyclic flats form an antichain. 
Hence, the largest size of a chain of nonempty proper cyclic flats of elementary split matroids is $h = 1$, and thus Theorem~\ref{thm:conn-alg-correct} implies the Algorithm~\ref{alg:conn} uses only $O(n)$ independence queries for this matroid class.

\smallskip

Finally, we remark that deciding matroid connectedness requires at least a linear number of independence queries, even for bounded error randomized algorithms.
\begin{proposition}\label{prop:conn-lower-bound}
    For any $n \ge 2$, deciding the connectedness of a matroid on a ground set with $n$ elements requires $\Omega(n)$ independence queries in the worst case. This holds for determininistic algorithms and randomized algorithms succeeding with probability at least $2/3$ on every input.
\end{proposition}
The proof of Proposition~\ref{prop:conn-lower-bound}, reported in Appendix~\ref{app:prop-conn-lb-proof} for brevity, uses the lower-bound construction of Huang et al.~\cite[Section~3]{HuangFengLi2026}.
Finally, Proposition~\ref{prop:conn-lower-bound} implies that Algorithm~\ref{alg:conn} is tight for direct sums of uniform matroids and elementary split matroids.

\medskip

\paragraph{Final remarks.} 
We established that matroid component decomposition has $\Theta(n\log n)$ independence query
complexity, even with bounded-error randomization, while matroid connectedness requires $O(n)$ queries for elementary split matroids, direct sums of uniform matroids and, more generally, matroid classes with bounded cyclic-flat chain length.
Whether this linear bound extends to arbitrary matroids remains open.
Interestingly, the worst-case query complexity of both tasks also remains open in the rank oracle model. In particular, a linear rank-query algorithm for component decomposition would answer the question of Chakrabarty and Liao~\cite{ChakrabartyLiao2024}
%\cite[Section~4]{ChakrabartyLiao2024} 
on learning general partition matroids with $O(n)$ rank queries.

\begin{credits}
\subsubsection{Acknowledgments.}
Manuel Aprile was partially supported by the SID project ``Network
optimization problems under extra constraints'' (BIRD 2024, University of
 Padua). % da inserire nella versione finale non blind
    \paragraph{AI disclosure.} ChatGPT (OpenAI) and Claude (Anthropic) were used to assist with language editing, refining the exposition, and improving as well as suggesting some mathematical arguments. The authors take full responsibility for the correctness and originality of all results.
\end{credits}

\bibliographystyle{splncs04}
\bibliography{references}

\newpage

\appendix

\section{Proof of Proposition~\ref{prop:counter}}\label{app:prop-counterex}

\begin{proof}
    Let $B^*=E_0-x+y$, where $x\in E_0$ and $y\in\overline{E}_0$.
    Since $r\ge 2$ and $n-r\ge 2$, there exist $x'\in E_0\setminus\{x\}$ and
    $y'\in\overline{E}_0\setminus\{y\}$.
    Consider the two sets $B_1=E_0-x'+y$ and $ B_2=E_0-x+y'$, which both belong to $\mathcal B\setminus\{B^*\}$.
    Since $x\in B_1\setminus B_2$ and $B_2\setminus B_1=\{x',y'\}$, the basis-exchange axiom requires that $B_1-x+z$ belong to $\mathcal B\setminus\{B^*\}$ for some $z\in\{x',y'\}$.
    However, $B_1-x+x'=B^*$, while
    \[
        B_1-x+y'=E_0-\{x,x'\}+\{y,y'\}
        \notin\mathcal B.
    \]
    Thus, the basis-exchange axiom is violated, implying that $\mathcal B\setminus\{B^*\}$ cannot be the family of bases of a matroid.
\end{proof}

\section{Simulating circuit-CUT queries with independence queries}
\label{app:corr-cut-simulation-proof}

\begin{corollary}\label{cor:cut-simulation}
Let $\mathcal{S}$ be a procedure that, given independence-oracle access to $M_\pi$ for an
unknown $\pi\in S_m$, answers %adaptively chosen 
CUT queries on $\mathcal{H}(M_\pi)$
exactly. Suppose that each query is answered using at most $t$ independence queries in
expectation, conditioned on all previous queries and answers. Then $t=\Omega(\log n)$,
where $n=2m$. %This holds even though $\mathcal{S}$ knows that the input belongs to the family $\{M_\pi\}$ and may share information across queries.
\end{corollary}

\begin{proof}
The circuits of $M_\pi$ are the pairs $\{b_i,x_{\pi(i)}\}$, so $\mathcal{H}(M_\pi)$ is a
perfect matching on $n$ vertices. Its components are the components of $M_\pi$.
By~\cite{ChakrabartyLiao2026}, there is a randomized algorithm $\mathcal{A}$ that
recovers the components of every hypergraph on $n$ vertices with probability at least
$2/3$, using at most $cn$ CUT queries in expectation, for a constant $c$.

Run $\mathcal{A}$ on $\mathcal{H}(M_\pi)$ and use $\mathcal{S}$ to answer each CUT query.
Let $Q$ be the number of CUT queries, and let $Z_k$ be the number of independence
queries used for the $k$-th CUT query. The event $\{Q\ge k\}$ is determined before the $k$-th
simulation starts. Hence the total number $T$ of independence queries satisfies
\[
\mathbb{E}[T]=\sum_{k\ge1}\mathbb{E}\bigl[Z_k\mathbf{1}_{\{Q\ge k\}}\bigr]
\le t\sum_{k\ge1}\Pr[Q\ge k]=t\,\mathbb{E}[Q]\le ctn .
\]
The composed algorithm recovers the components of every $M_\pi$ with probability at least
$2/3$. By Theorem~\ref{thm:decompose-lower-bound}, some $\pi$ satisfies
$\mathbb{E}[T]\ge(\log_2(m!)-3)/6$. Therefore
$t\ge(\log_2(m!)-3)/(6cn)=\Omega(\log n)$.

\end{proof}

\section{Proof of Corollary~\ref{cor:uni-sum-lin-cmp}}\label{app:cor-uni-sum-proof}
\begin{proof}
    The correctness of Algorithm~\ref{alg:conn} for $M$ follows from Theorem~\ref{thm:conn-alg-correct}. We now show that the algorithm requires a linear number of independence queries.
    
    By hypothesis, $M$ has the form $\bigoplus_{i = 1}^t U_{r_i, |E_i|}$
    where, for each $i \in [1, t]$, $r_i$ is a nonnegative integer, and the sets $E_i$ are pairwise disjoint.
    For the basis $B$ constructed by Algorithm~\ref{alg:conn}, $B_i = B \cap E_i$ is the basis of the $i$-th summand of $M$.
    For each $e \in E_i \setminus B_i$, the fundamental circuit $C(e, B)$ is $B_i + e$, and thus $N_B(e) = B_i$.

    If $r_i$ is positive for each $i \in [1, t]$ and $M$ has at least a summand distinct from $U_{1, 1}$, then Algorithm~\ref{alg:conn} does not return \textsc{Disconnected} at line~\ref{alg:conn:ret-disc-1} or line~\ref{alg:conn:ret-disc-2}.
    Next, the algorithm chooses an element $e_0 \in E_j$ for some $j \in [1, t]$, and thus line~\ref{alg:conn:init-S} initializes $S$ to $N_B(e_0) = B_j$.
    If $N_B(e) \cap S \ne \varnothing$, then $e \in E_j$ because the summands of $M$ are disjoint. Hence, $N_B(e) = B_j = S$ holds, and thus $N_B(e) \setminus S = \varnothing$, implying that the conditions at lines~\ref{alg:conn:c1} and \ref{alg:conn:c2} cannot hold simulteneously for $M$.

    Observe that $S = B$ if and only if $t = 1$; in this case, Algorithm~\ref{alg:conn} does not enter the \text{while} loop and correctly returns \textsc{Connected} at line~\ref{alg:conn:ret-conn}.
    Now, suppose that $t \ge 2$. In this case, Algorithm~\ref{alg:conn} enters lines~\ref{alg:conn:innermostfor-begin}--\ref{alg:conn:innermostfor-end}. By the argument above, Algorithm~\ref{alg:conn} does not find any $f \in E \setminus B$ satisfying the conditions at lines~\ref{alg:conn:c1} and \ref{alg:conn:c2}, and thus returns \textsc{Disconnected} at line~\ref{alg:conn:ret-disc-3}.

    Hence, lines~\ref{alg:conn:first-B}--\ref{alg:conn:init-S} use $O(n)$ independence queries, while lines~\ref{alg:conn:while-begin}--\ref{alg:conn:while-end} use at most $2(n - r)$ independence queries, and thus Algorithm~\ref{alg:conn} uses $O(n)$ independence queries to decide the connectedness of $M$.
\end{proof}

\section{Proof of Proposition~\ref{prop:conn-lower-bound}}\label{app:prop-conn-lb-proof}

\begin{proof}
    Let $E$ be a ground set of $n$ elements. Let $M$ be the uniform matroid $U_{1,n}$ on $E$ and, for each $e\in E$, let $M_e=U_{1,n-1}\oplus U_{0,1}$, where the rank-zero summand has ground set $\{e\}$. The matroid $M$ is connected, while $M_e$ is disconnected. Their independence oracles differ only on the query $\{e\}$.

    Consider an algorithm with error probability at most $\varepsilon$ on every input. For each $e\in E$, compare its runs on $M$ and $M_e$ using the same random bits. %The runs proceed identically until the two algorithms query $\{e\}$.
    By the union bound, both runs are correct with probability at least $1-2\varepsilon$. On this event their outputs differ, so in particular the run on $M$ must query $\{e\}$. Hence, the probability that the run on $M$ queries $\{e\}$ is at least $1-2\varepsilon$, and this hold for any $e\in E$.

    The total number $Q$ of queries on $M$ is at least the number of
    distinct singletons queried. Consider the expectation of $Q$ over the random bits.
    By linearity of expectation, $\mathbb{E}[Q]$ is lower bounded by the sum, for each $e \in E$, of the probability that $\{e\}$ is queried on $M$; thus we obtain \[
    \mathbb{E}[Q] \geq n(1-2\varepsilon).
    \] %this sum, in turn, is not smaller than $$.

    \noindent
    Taking $\varepsilon=0$ and $\varepsilon=1/3$ gives the deterministic and randomized bounds, respectively.
\end{proof}

\end{document}